# Going green across boundaries: Spatial effects of environmental policies on tourism flows[*]

Riccardo Gianluigi Serio[1], Diego Giuliani[1], Maria Michela Dickson[2], and Giuseppe Espa[1]

[1]Department of Economics and Management, University of Trento, Italy

[2]Department of Statistical Sciences, University of Padova, Italy

## Abstract

This study investigates the relationship between environmental sustainability policies and tourism flows across Italian provinces using a Spatial Durbin Error Model (SDEM) within a gravity framework. By incorporating both public and corporate environmental initiatives, the analysis highlights the direct and spatial spillover effects of sustainability measures on tourism demand. The findings indicate that corporate-led initiatives, such as ecocertifications and green investments, exert a stronger direct influence on tourism flows compared to public measures, underscoring the visibility and immediate impact of private sector actions. However, both types of initiatives generate significant positive spatial spillovers, suggesting that sustainability efforts extend beyond local boundaries. These results demonstrate the interconnected nature of regional tourism systems and emphasize the critical role of coordinated sustainability policies in fostering tourism growth while promoting environmental protection. By addressing the spatial interdependencies of tourism flows and sustainability practices, this research provides valuable insights for policymakers and stakeholders seeking to improve sustainable tourism development at regional and national levels.

[*]Corresponding author: Riccardo Gianluigi Serio. Email: riccardo.serio@unitn.it

# 1 Introduction

Tourism has become a central economic force in modern societies, profoundly influencing lifestyles as disposable income and leisure time increase (Cracolici and Nijkamp, 2008). Beyond serving as a catalyst for economic growth (Lee and Brahmasrene, 2013), the tourism sector significantly increases global GDP, contributing 10.4% in 2019 (WTTC, 2021). For policymakers, economic stakeholders, and decision makers aiming to optimize resources and forecast trends in tourism, understanding the determinants of tourism flows is crucial. Early studies focused primarily on demand-side variables to explain crossborder tourism, focusing on factors such as income levels, population size, relative prices, and geographical distance, which often approximates transportation costs (Lim, 1997). Although valuable, this traditional approach may not capture the evolving preferences of tourists and the growing variety of destinations catering to specific interests and niche markets (Papatheodorou, 2001). This evolution highlights the need for a multidimensional analytical framework that incorporates both demand and supply factors to fully grasp the complexities of modern tourism dynamics. As tourist preferences have evolved, increasingly specific factors now influence destination choices. Research on the success of tourist destinations has identified critical components, such as cultural, historical and environmental elements, as part of an integrated model in which destinations offer a rich blend of products and services (see, among others, Crouch and Ritchie, 1999; Buhalis, 2000; Ritchie and Crouch, 2000; Enright and Newton, 2004; Ruhanen, 2007; Cracolici and Nijkamp, 2008). Within this framework, local stakeholders strive to develop cohesive offerings that capitalize on regional assets, thus improving the competitive edge of the destination in a diverse global tourism market (Teece et al., 1997).

In addition, environmental sustainability has become increasingly integral to tourism development, recognized as essential by policymakers and industry stakeholders alike. Incorporating sustainability into tourism policy began in the late 1980s, aligning with the larger embrace of principles of sustainable development (Hall, 2011). Since then, national governments and international organizations have actively crafted frameworks to guide the ecological transformation of the tourism sector. Despite these initiatives, there is uncertainty about the actual impact of these policies. Many researchers (see, among others, Guo et al., 2019; Torres-Delgado and Palomeque, 2012) continue to highlight the need for more rigorous analysis to evaluate and improve the effectiveness of sustainable tourism strategies. In addition, some others (Farsari et al., 2007) have proposed theoretical frameworks to differentiate between policies that merely aim to improve the sustainable image of a destination and those that are genuinely committed to environmental preservation, a strategy found to be more effective in reducing the ecological footprint of the tourism sector.

The purpose of this study is to investigate the relationship between sustainable adoption and tourism demand in Italian provinces. A key innovation of this research is the inclusion of environmental commitment variables to assess the relationship between sustainable destination practices and tourist attraction. Since previous research (e.g., Serio et al., 2024) found a positive association between tourism demand and the level of environmentally sustainable infrastructure, our objective is to investigate whether proximity between municipalities improves the diffusion of sustainable investments in neighboring areas, potentially establishing a virtuous cycle of environmental protection and economic prosperity driven by tourism activities. Spillover effects are common in tourism flows, referring to the indirect or unintended impacts that a region's tourism industry may have on tourism flows to other regions (see, among others, Drakos and Kutan, 2003; Neumayer, 2004; Gooroochurn and Hanley, 2005). This is especially true for long-haul tourists, who are motivated to maximize the utility of their travels by exploring multiple nearby destinations. Such travel patterns inherently strengthen the connectivity of tourism flows across regions. This interconnectedness underscores the importance of accounting for spatial dependence in tourism flow analyses. The patterns of tourist movement are shaped not only by the immediate appeal of the origin and destination but also by the attractions and characteristics of the surrounding areas. This complexity requires a comprehensive analytical approach that recognizes multi-regional influences to accurately capture the dynamics of tourism flows (Curry, 1972). By incorporating environmental commitment variables and accounting for spatial spillovers through a gravity model integrated with a Spatial Durbin Error Model, we empirically assess the relevance of proximity across Italian provinces in shaping and spreading innovation and investments in green adoption, both in the private and public sectors, related to the tourism industry. This approach enables us to estimate how the effects of the internal determinants of the provinces are amplified by the influence of neighboring areas and to evaluate the impact of sustainable investments on tourist attraction, including the potential contagion effect of sustainable practices among neighboring provinces. The paper is structured as follows. Section 2 reports a review of the literature on tourism demand models and applications of environmental policies. Section 3 explains the gravity model with spatial expansion used on the data presented in Section 4, in which the results of four models are presented and commented on. Section 5 concludes the paper.

## 2 Literature background

Tourism demand has been analyzed in literature with gravity models, particularly in contexts where bilateral tourist flows depend on both economic and geographic factors. Traditional models emphasize the role of income, population, distance, and price variables, creating a foundation to assess tourism inflows between origin-destination pairs, where demand is proportional to economic 'mass' (GDP and population) and is inversely related to distance. The gravity model, initially used in international trade, has theoretical roots in various economic models. Colwell (1982) connected tourism demand to traditional gravity models through maximization of utility, while Morley et al. (2014a) strengthened the link of the model to individual utility. Recent work by Santana-Gallego and Paniagua (2022) further tailored the gravity model for tourism by adapting Anderson (2011) migration model. They introduced determinants such as the utility of tourism satisfaction and travel costs, highlighting factors such as bilateral agreements, GDP of origin and destination, and destination-specific attributes (e.g., coastlines, events). This modified framework reflects the aggregate tourism demand based on determinants of time, origin, and destination. In empirical applications, as Morley et al. (2014b) points out, studies can be based on a time series (limiting the number of destinations) or multiple originsdestination data perspective (deflecting attention from time-related determinants to space-related ones). Despite the interest of scholars being mainly focused on both origin and destination determinants, some works focus the attention on push factors in explaining most of the variability in tourism flows (such as the GDP in the origin, Garín-Muñoz, 2009), some others (e.g., Zhang and Jensen, 2007) center the attention on pull determinants, arguing that most of tourism demand can be mainly explained through destination's characteristics.

Notwithstanding its effectiveness in studies related to tourism demand, scholars have recognized the need to enhance the gravity model to capture additional nuances, particularly those arising from spatial dependencies, as regions are rarely isolated in factors influencing their attractiveness to tourists (see, among others, Curry, 1972; Mata and Llano-Verduras, 2012). Traditional gravity models often overlook the influence of spatial interactions between regions. In tourism, spatial dependencies mean that the attractiveness of a destination is not isolated but is affected by the characteristics and policies of its neighbors. Marrocu and Paci (2013) advanced this perspective by incorporating spatial econometric techniques into gravity models to account for spillover effects across provinces. In fact, tourists often consider nearby regions as complementary destinations, creating spatial interdependencies. Using spatial interaction models of origin-destination, Marrocu and Paci (2013) demonstrated that tourism flows are not only determined by bilateral factors. Instead, spatial lags and the characteristics of neighboring regions have a significant effect on both inbound and outbound tourist flows. This supports the idea that tourists share recommendations and preferences in neighboring origin areas, creating spillovers on the demand side and the supply side. From the demand side, tourists from neighboring provinces of origin may have similar cultural and socioeconomic backgrounds, influencing their destination preferences. On the supply side, contiguous destinations often share infrastructure and resources, such as airports, which facilitate tourism inflows.

Due to its large worldwide economic impact, tourism is also set as a major contributor to environmental degradation. Recent literature has struggled to offer a deeper understanding of the complex environmental dynamics affecting the tourism industry. For example, Gössling et al. (2012) focuses his attention on carbon dioxide emissions (considered an overall proxy to measure the environmental footprint of regional tourism activity). As reported by Tsionas and Assaf (2014), several scholars focus their efforts on offering measures of eco-efficiency of destinations. This approach may underestimate the role of evolution and improvement achieved by a destination (in terms of environmental efficiency) over time, due to the stasis of the measurement. However, tracking the evolution of environmental adoption, specifically for the tourism sector, is no simple task due to various factors such as the lack of univocal standards and the rigorous data follow-up record (Mycoo, 2014), thus making the study of the phenomenon from a temporal perspective increasingly challenging.

Regarding institutional efforts to provide measures and frameworks, since 1992 the European Union (EU) and other public bodies have pioneered efforts to encourage sustainable tourism practices by establishing the EU Ecolabel, the Eco Management and Audit Scheme (EMAS), the Blue Flag (BF) certification, and the Airport Carbon Accreditation (ACA) program. In addition to public schemes, private initiatives have emerged within the tourism sector to reduce environmental impact, such as the BioHotels label and the Green Key certification. Although these certifications reflect an increase in environmental awareness, there is still uncertainty regarding the effect of such initiatives on tourism flows and regional economic growth. Research has highlighted several limitations in sustainable tourism policies. Scholars argue that a primary issue is the absence of a clear definition of "tourism policy", leading to vague and adaptable policy frameworks (see, among others, Guo et al., 2019; Farsari et al., 2007). Many sustainable tourism policies prioritize economic goals over environmental sustainability (Yüksel et al., 2012), which complicates their practical application and limits the scope of post-implementation evaluation. Furthermore, the evaluation of these policies is often hindered by insufficient and fragmented data collection and complex guidelines that intertwine with broader policies (Dodds and Butler, 2009). Although governments play an essential role in guiding destinations towards sustainable practices (Guo et al., 2019), they sometimes lack the consistent leadership needed for the ecological transition (Andersen et al., 2018).

## 3 Gravity for tourism flows and model spatial expansion

The idea behind the gravity model is linked to the universal gravitational law of Isaac Newton, which states that the strength of a bond between two entities is directly proportional to their masses and inversely proportional to their distance. Transposed and adapted to the tourism flow, we can assume that the strength of a tourism phenomenon (in terms of number of visits) between an origin and a destination can be studied and modeled considering the masses (e.g. GDP, population) and resistances (e.g. distance). For simplicity, we present the formal equation (using matrix notation) for the gravity model proposed by Morley et al. (2014a), which is considered the most advanced and recent formulation since the initial attempt made by Zipf (1946).

The tourism flows from $m$ origin countries to $n$ destination regions are organized in the $n$ by $m$ matrix $NV$. The dependent variable $NV$ used in the gravity model is derived by applying the matrix operator $vec$, which stacks the columns of the matrix $NV$ in a single column vector: $NV = vec(NV)$. Consequently, $NV$ consists of $N = n \times m$ observations: the first $n$ observations correspond to the first origin country, the next $n$ observations correspond to the second origin country, and so on. The determinants of tourism flows originating from each country - that is, the "pushing" factors that influence outbound tourism (e.g., GDP, accessibility to transport, population) - are represented in the $m$ by $s$ matrix $X_O$, where $s$ denotes the number of explanatory variables specific to the origin. Similarly, the "pulling" factors that attract visitors to each destination such as GDP, population, cultural and natural attractions, sustainable infrastructure, and environmental policies - are captured in the $n$ by $p$ matrix $X_D$, where $p$ is the number of explanatory variables specific to the destination. Furthermore, the columns of the $N$ by $r$ matrix $OD$ represent the characteristics of the specific origin-destination (e.g., the geographic distance between the origin and destination).

To align $X_O$ with $NV$, it is reshaped into the $N$ by $s$ matrix $O$, defined as $O = X_O \otimes \iota_n$, where $\iota_n$ is a column vector of ones. The Kronecker product operator, $\otimes$, ensures that each row of $X_O$ is repeated $n$ times, thus linking the $s$ origin-specific characteristics to the destination. Similarly, $X_D$ is transformed into the $N$ by $p$ matrix $D$ as $D = \iota_m \otimes XD$, with its columns capturing the $p$ destination-specific attributes for each origin. The gravity model is then specified as follows:

$$nv = \alpha_{\iota N} + O\beta_O + D\beta_D + OD\beta_{OD} + \varepsilon. \tag{1}$$

Here, the dependent variable $nv$ is the natural logarithm of $NV$, and $\alpha_{\iota N}$ represents the intercept. As described above, the matrices $O$, $D$, and $OD$ contain information on the characteristics of the origin countries, the attributes of the destination regions, and the interaction variables between origins and destinations, respectively. The corresponding effects are captured by the coefficient vectors $\beta_O$, $\beta_D$, and $\beta_{OD}$. The error term, $\varepsilon$, is assumed to be normally, independently and identically distributed, consistent with the standard assumption in gravity models that origin-destination flow observations are independent once the influence of distance has been accounted for. To reduce disturbance arising from omitted variable issues and unobserved heterogeneity, Kuminoff et al. (2010) suggest that the inclusion of fixed effects should be considered the preferable approach to reduce spatial heterogeneity in cross-sectional data. To avoid overestimation of the effects by excluding spatial dependence issues, spatial proximity among units has been integrated within the model. In this way, employing the province-province spatial interaction models (see, among others, Marrocu and Paci, 2013), we can split the total effect of a variable into direct and indirect effects, which allows us not to exclude the main spatial dependencies between the host provinces. Since our data capture the unilateral flow of inbound tourism, we do not consider bilateral flows (Marrocu and Paci, 2013), but we only account for spatial interactions between destinations.

In line with Marrocu and Paci (2013), we employ a spatial econometric framework to incorporate the system of spatial connections between regions. These connections are defined using a spatial weight matrix, $W(w_{ij})$, which is an $m$ by $m$ non-negative matrix where each element $w_{ij}$ represents the degree of proximity between region $i$ and region $j$. The measure of proximity is based on the critical cutoff neighborhood criterion, expressed as

$$w_{ij} = \begin{cases} 1 & if\, d_{ij} \leq d_c\, and\, i \neq j \\ 0 & otherwise \end{cases}$$

where $d_{ij}$ is the geographical distance between the centroids of the regions $i$ and $j$, and $d_c$ is the critical threshold, set at 120 kilometers in our application. To facilitate interpretation and align with common practice, the spatial weight matrix $W$ is row-standardized prior to model estimation.

The gravity model presented in Equation 1 can be extended to account for spatial dependence between regions and the potential presence of spatial spillovers by adopting the *Spatial Durbin Error Model* (SDEM):

$$nv = \alpha \iota_N + O\beta_O + D\beta_D + OD\beta_{OD} + W_D D\theta_D + W_D OD\theta_{OD} + u \quad (2)$$
$$u = \lambda W_D u + \varepsilon.$$

In this formulation, $W_D = I_m \otimes W$, where $I_m$ is the $m$ by $m$ identity matrix. The spatially lagged terms $W_D D$ and $W_D OD$ represent the averages of the destination-specific and origin-destination-specific independent variables for neighboring regions. The regression parameters $\theta_D$ and $\theta_{OD}$ are estimates of (local) spatial spillover effects as they describe the impact on tourism flows to a region arising from changes in the characteristics of nearby regions. Finally, the spatial autoregressive coefficient $\lambda$ is a parameter that needs to be estimated to control for the potential dependence in the model disturbances. The SDEM is chosen to capture local spatial spillovers rather than global ones because tourism is inherently a local phenomenon. Models for global spillovers, like the spatial Durbin model or the spatial autoregressive model, are less suitable in this circumstance. It is indeed unlikely that changes in a region's characteristics would ripple through the entire national territory via neighboring regions.

## 4 Empirical application

### 4.1 Data description

The data described here are collected by the Italian National Institute of Statistics (ISTAT), which tracks flows from global origins (NUTS 1) to Italian destinations (NUTS 3). The dependent variable is the count of overnight stays in all types of accommodation in $n = 107$ Italian provinces for the year 2019. To align with the study focus on environmental policies in tourism, the analysis is limited to $m = 32$ European countries, creating a consistent legislative and socioeconomic context between origins and destinations. Distance is a key independent variable in tourism studies, often represented as both geographical and travel distance. Generally, the expected effect of distance on tourism flows is negative, meaning that as distance increases, the utility and thus the tourism demand decrease. For this study, distances are calculated in Cartesian space, and additional variables include GDP per capita, for both origins and destinations, which reflects economic capability and travel propensity due to higher income levels. To address population influences, we include the population density at the destination, anticipated to have a negative association with tourism flows, consistent with the findings on domestic Italian tourism (Marrocu and Paci, 2013). Additional factor includes destination unemployment rates to capture production levels. Geographical and accessibility characteristics, such as infrastructure levels, air travel options, and shared borders, are frequently considered in tourism studies (Roselló Nadal and Santana

Gallego, 2022). Note that 69 of Italian 107 provinces lack air services. This study includes an accessibility metric obtained from Espon, covering all transport modes, where higher accessibility is expected to positively impact tourism flows by easing transport and travel logistics. Environmental and climate variables also play a role in the tourism literature (Roselló Nadal and Santana Gallego, 2022). Previous studies (Marrocu and Paci, 2013) show that Blue Flag certifications for coastal areas positively impact tourism demand. Environmental variables have varied impacts depending on region; for instance, southern Italian destinations tend to attract tourism for environmental reasons, while northern regions attract visitors for cultural activities (Massidda and Etzo, 2012). Furthermore, research on air pollution in China finds that levels of pollution at both the origin and destination can reduce the arrivals of tourists (Xu and Dong, 2020). For this study, environmental variables are organized into two categories to evaluate the effect of sustainability policies on tourism flows.

The first category, namely *Public sustainabiliy measures*, includes green certifications and sustainability policies led by local governments or public entities. Indicators here are the number of Blue Flags and EMAS-labeled public bodies and waste management firms within each province. The second category, *Corporate sustainability measures*, tracks sustainable practices in tourism-related businesses, covering EMAS-accredited accommodations, Ecolabel-certified entities, Airport Carbon Accreditation, Bio Hotels, and Green Key projects within each province. Variables related to origin countries are labeled with $O$, while destination variables are labeled with $D$. Spatial lagged variables are indicated with $W_D$.

We run four models of increasing complexity, starting from a basic model with two masses (GDP and distance).

## 4.2 Results

The SDEM implementation provides an interesting analysis of tourism flows, capturing both direct effects of key variables and spatial spillovers across neighboring provinces. The results align with the foundational principles of gravity models, while also revealing new insights into the role of sustainability measures, socioeconomic factors, and environmental quality in shaping the dynamics of tourism. Table 1 shows the results of the SDEM estimation through incremental enhancements of the model.

|  | Model 1 | Model 2 | Model 3 | Model 4 |
|---|---|---|---|---|
| (Intercept) | $-14.22^{***}(4.30)$ | $-12.96^{*}(7.35)$ | $6.44(7.68)$ | $56.02^{***}(8.72)$ |
| Distance | 0.54(0.54) | −0.94***(0.34) | −1.14***(0.33) | −1.42***(0.32) |
| GDP$_O$ | 0.47***(0.16) | 0.59***(0.16) | 0.57***(0.15) | 0.56***(0.15) |
| GDP$_D$ | 3.77***(0.16) | −0.19(0.16) | −0.46***(0.17) | −0.31**(0.14) |
| Unemployment$_D$ | — | −2.01**(0.83) | −2.35***(0.83) | −2.71***(0.72) |
| Density$_D$ | — | −0.16***(0.04) | −0.05(0.04) | −0.02(0.03) |
| Accessibility$_D$ | — | 0.08(0.12) | 0.09(0.13) | −0.22**(0.11) |
| Museum visitors$_D$ | — | 0.40***(0.03) | 0.47***(0.03) | 0.54***(0.02) |
| Coasts$_D$ (km) | — | 0.12***(0.01) | 0.15***(0.01) | 0.17***(0.01) |
| Domestic flows | — | 5.01***(0.65) | 4.96***(0.60) | 2.48***(0.34) |
| No. tourism firms$_D$ | — | 1.20***(0.04) | 1.03***(0.04) | 0.92***(0.04) |
| Air pollution$_D$ | — | −1.18***(0.10) | −1.13***(0.10) | −1.07***(0.09) |
| Corporate sustainability measures$_D$ | — | — | 0.08(0.08) | 0.26***(0.07) |
| Public sustainability measures$_D$ | — | — | 0.22***(0.04) | 0.17***(0.03) |
| $W_D$Distance | −1.28**(0.55) | 0.59(0.37) | 0.75**(0.36) | 0.83**(0.34) |
| $W_D$GDP$_D$ | −1.79***(0.35) | 1.99***(0.62) | 1.00(0.63) | 2.03***(0.39) |
| $W_D$Unemployment$_D$ | — | 0.12(3.07) | 5.77*(3.08) | −0.60(1.73) |
| $W_D$Density$_D$ | — | 0.44***(0.16) | 1.00***(0.17) | 1.10***(0.11) |
| $W_D$Accessibility$_D$ | — | −0.91**(0.39) | −0.60(0.39) | −1.91***(0.27) |
| $W_D$Museum visitors$_D$ | — | −0.16*(0.10) | 0.00(0.10) | 0.36***(0.07) |
| $W_D$Coasts$_D$ | — | −0.34***(0.05) | −0.21***(0.05) | 0.06**(0.03) |
| $W_D$No. tourism firms$_D$ | — | 0.06(0.14) | −1.26***(0.17) | −1.92***(0.09) |
| $W_D$Air pollution$_D$ | — | −1.48***(0.43) | −2.81***(0.43) | −3.18***(0.33) |
| $W_D$Corporate sustainability measures$_D$ | — | — | 2.39***(0.43) | 3.58***(0.32) |
| $W_D$Public sustainability measures$_D$ | — | — | 1.58***(0.16) | 1.41***(0.13) |
| $\lambda$ | 0.74***(0.01) | 0.84***(0.01) | 0.83***(0.01) | 0.29***(0.03) |
| Num. obs. | 3424 | 3424 | 3424 | 3424 |
| Parameters | 8 | 23 | 27 | 56 |
| Log Likelihood | −6522.50 | −4995.63 | −4919.98 | −4567.91 |
| AIC (Linear model) | 15061.43 | 13281.94 | 13129.65 | 9344.82 |
| AIC (Spatial model) | 13060.99 | 10037.26 | 9893.96 | 9247.83 |
| LR test: statistic | 2002.43 | 3246.69 | 3237.69 | 98.99 |
| LR test: p-value | 0.00 | 0.00 | 0.00 | 0.00 |

***$p < 0.01$; **$p < 0.05$; *$p < 0.1$

**Table 1:** Maximum likelihood estimates of SDEM for the determinants of tourism flows. Variables are taken in logarithmic form (except for the dummy for *Domestic flows* and the two count measures of sustainable initiatives). Values in parentheses are the standard errors. (*Significant at $p < 0.1$, **$p < 0.05$, ***$p < 0.01$.)

We start by noting that the spatial autoregressive parameter $\lambda$ is positive and significant across all models, confirming strong spatial dependencies in tourism flows. This indicates that tourism in one province is influenced by flows in neighboring provinces, highlighting the interconnectedness of regional tourism dynamics. The basic specification of the gravity model (Model 1) confirms the established findings in the literature. As expected, the distance coefficient is negative and significant in all other models, reinforcing the fundamental premise that tourism flows decrease as the distance between the origin country and the destination province increases. The elasticity in Model 4 is slightly larger in magnitude than those reported in previous studies (see e.g., Marrocu and Paci, 2013), which often find elasticities closer to -1, which potentially reflects the inclusion of domestic flows in this analysis. However, the spatial lag of distance offers additional nuance. In Model 1 the coefficient is negative, suggesting that greater distances from neighboring provinces act as a deterrent to tourism flows in the focal province. However, in Models 3 and 4, the lagged distance effect becomes positive and statistically significant. This shift reflects a complementary effect among geographically proximate provinces, where shorter distances from the origin to neighboring provinces can set opportunities for multi-destination trips, where tourists visiting one province are likely to explore nearby provinces as well, thereby reducing the effective resistance of distance. These results highlight the dual role of distance. Although greater distances generally reduce tourism flows, proximity to neighboring provinces can offset this resistance and enhance the attractiveness of regional tourism. In other words, the inclusion of spatial lags modifies and moves one step further the interpretation of this relationship, as neighboring provinces appear to mitigate the effect of distance. The spatial lag of distance suggests that provinces closer to the origin exert a complementary pull effect on the focal province.

The role of GDP further reinforces the conceptual underpinnings of the gravity model. The GDP of the origin has a consistently positive and significant effect, confirming that the wealthier countries of origin generate more substantial tourism flows. This finding is consistent with the notion that tourism is a normal good, where higher disposable incomes at the origin allow more frequent or longer distance travel (Rosselló Nadal and Santana Gallego, 2022). In contrast, the GDP of the destination province has a negative direct effect on the final model. This result is somewhat counterintuitive but may reflect the diminishing returns of economic development in wealthier provinces, where tourism infrastructure and demand may already be saturated. Alternatively, it may signal a price effect, where wealthier provinces are perceived as more expensive and thus less attractive to cost-sensitive tourists. Interestingly, also in this case the lagged results for the destination's GDP mitigate the direct effect. In fact, the spatial delay of GDP for destination provinces is positive and highly significant, indicating that wealth in neighboring provinces positively influences tourism flows to the focal province, possibly through shared regional branding or complementary attractions.

Among socioeconomic variables, unemployment in the destination province exhibits a significant negative effect, suggesting that higher unemployment rates reduce the perceived attractiveness of a province. This may be due to the fact that the tourism sector is a relevant job generator in Italy, so higher unemployment rates are registered in Italian provinces less involved in tourism activities. In addition, negative perceptions of safety, service quality, or overall economic vitality can be positively correlated with unemployment and negatively with tourism attraction. The number of tourism firms positively influences the tourism flows, with an elasticity of 0.92 in the preferred model.

This finding reflects the role of tourism infrastructure in improving the appeal of a destination, as a higher concentration of tourism-related businesses signals better services and facilities for visitors. However, the spatial delay of the tourism companies has a negative effect, suggesting the presence of strong competition between neighboring provinces that can tend to dilute the benefits of local tourism infrastructure. This result suggests that, while local investments in tourism infrastructure are beneficial, their impact may be limited in regions with high competition. Accessibility, which represents the degree of efficiency of the provincial transport infrastructure, has a mixed and unexpected impact. Although initially positive in Model 2 and 3 (but with low significance), its direct effect becomes negative in the final specification, possibly indicating that although the transport infrastructure is more developed in northern Italy, this does not necessarily come with a higher inbound tourism activity, since also southern Italy experiences a strong tourism activity, as well as lower levels, on average, of economic development and accessibility infrastructures. Domestic flows also play a crucial role, with a strong positive effect. This finding underscores the importance of domestic tourism as a stable and significant contributor to overall tourism demand, particularly in the context of regional travel. Cultural and natural amenities remain critical drivers of tourism flows, consistent with established literature, e.g. Massidda and Etzo (2012). The number of visitors who have access to museums has a robust positive effect in Model 4, highlighting the enduring appeal of cultural heritage. The spatial lag of museum visitors further illustrates the complementary nature of cultural attractions, with a positive and significant elasticity. This suggests that provinces with strong cultural offerings enhance the attractiveness of neighboring provinces, possibly due to multi-destination travel behavior thus setting the roots for the development of macro cultural hubs. The length of the coast also exerts a positive and significant influence, underscoring the role of natural landscapes, particularly coastal areas, in the attraction of tourists. In contrast, compared to museum visitors, the spatial lag of coastal length transitions from negative Models 2 and 3 to slightly positive in the final specification. This shift reflects a complex interplay between competition and collaboration between coastal provinces. Although initially competitive, coordinated marketing or complementary offerings among coastal regions can mitigate these effects, enhancing the overall attractiveness of the area. Air pollution has a consistently negative and significant impact on tourism flows. This finding highlights the deterrent effect of poor air quality on tourism demand, in agreement with previous studies that emphasize the importance of environmental aesthetics and health concerns in tourist decision making Xu and Dong (e.g., 2020). The spatial lag of this variable is even more negative indicating that pollution in neighboring provinces also significantly reduces tourism flows to the focal province. These results underscore the need for coordinated environmental policies in regions to address pollution and enhance tourism attractiveness.

In relation to this, a central contribution of this study lies in examining the impact of environmental sustainability measures on tourism flows. Corporate sustainability measures, such as eco-certifications for accommodations, emerge as a significant determinant. This finding underscores the importance of initiatives from the private sector in attracting environmentally conscious tourists. Public sustainability measures, including initiatives such as Blue Flag certifications and EMAS compliance, also have a positive and significant effect. The spatial spillovers of the sustainability measures provide additional insights into their regional dynamics. The spatial lag of corporate sustainability measures exhibits a large and significant coefficient. This indicates that sustainable practices in neighboring provinces amplify tourism flows to the focal province, possibly through regional branding effects, where tourists perceive the entire region as environmentally progressive. The spatial lag of public measures, while smaller in magnitude, remains significant. These findings suggest that the benefits of sustainability initiatives extend beyond individual provinces, emphasizing the importance of regional cooperation in the implementation of environmental policies and also opening the gates for the creation of specialized regional environmental hubs that account for the highest score in environmental investments made by companies operating within provincial boundaries. The findings for environmental variables reveal the critical role of sustainability in shaping tourism flows. Both corporate and public sustainability measures positively influence tourist demand, with corporate initiatives showing a stronger direct impact, likely due to their visibility and direct engagement with visitors. In particular, the significant spatial spillovers of both types of measures emphasize that sustainability efforts extend their influence beyond local boundaries, also benefiting neighboring provinces. This underscores the interconnected nature of regional tourism systems and highlights the importance of coordinated sustainability policies. Furthermore, the negative impact of air pollution, both locally and through spatial spillovers, reinforces the need for environmental quality to remain a priority in tourism planning. Together, these findings show that environmental sustainability is not only an ethical imperative, but also a practical strategy to enhance regional tourism competitiveness.

## 5 Conclusion

This study highlights the complex and interconnected relationship between environmental sustainability and tourism flows across Italian provinces, leveraging the Spatial Durbin Error Model within a gravity framework. By incorporating public and corporate environmental initiatives, the analysis uncovers not only their direct effects on tourism demand but also significant spatial spillovers. These findings reveal how sustainability efforts can influence not only the province that implements them but also neighboring regions, offering valuable information for tourism policy and regional planning. Our results confirm that both public and corporate environmental measures are positively related to tourism flows, but the effects of corporate initiatives stand out as particularly strong. Certifications and green practices adopted by private companies have a more immediate and visible impact on tourists, making these efforts highly influential in driving environmentally conscious travel behavior. Public sustainability measures, while also important, tend to have a more modest direct effect, likely reflecting their broader focus on long-term regional sustainability rather than direct consumer engagement. This distinction underscores the importance of the participation of the private sector in sustainability efforts, while suggesting that public measures play a complementary and enabling role. Perhaps even more compelling are the spatial spillovers observed for both types of sustainability measures. Corporate initiatives in neighboring provinces significantly improve tourism flows to a given province, indicating that regional branding and the perception of a broader environmentally friendly area can be powerful drivers of tourism demand. The results emphasize that sustainability in tourism is not just a local issue but is inherently regional. Tourists often perceive and respond to destinations as part of interconnected systems, making coordinated efforts across provinces a critical strategy to maximize the impact of sustainable environmental investments.